\documentclass[12pt]{article}
\usepackage{fullpage}
\usepackage{epsf}
\usepackage{epsfig}
\usepackage{amsthm}
\usepackage{amsfonts}
\usepackage{color}
\usepackage{amsmath}
\usepackage{setspace}
\usepackage{natbib}
\usepackage{url}
\usepackage{algorithmic}
\usepackage[boxed]{algorithm}
\usepackage[thinlines]{easymat}

\begin{document}

\newtheorem{Definition}{Definition}
\newcommand{\mathbbm}[1]{\mathbb{#1}}
\newcommand{\bm}[1]{\mbox{\boldmath $#1$}}
\newcommand{\mb}[1]{\mathbf{#1}}
\newcommand{\bE}[0]{\mathbb{E}}
\newcommand{\bP}[0]{\mathbb{P}}
\newcommand{\ve}[0]{\varepsilon}
\newcommand{\Var}[0]{\mathbb{V}\mathrm{ar}}
\newcommand{\Corr}[0]{\mathbb{C}\mathrm{orr}}
\newcommand{\Cov}[0]{\mathbb{C}\mathrm{ov}}
\newcommand{\mN}[0]{\mathcal{N}}
\newcommand{\iidsim}[0]{\stackrel{\mathrm{iid}}{\sim}}
\newcommand{\NA}[0]{{\tt NA}}
\newcommand{\argmax}{\operatornamewithlimits{argmax}}
\newcommand{\acf}{ACF}
\newcommand{\iid}{iid}
\newcommand{\backshift}{\ensuremath{\mathcal{B}}}
\newcommand{\innov}{\ensuremath{\varepsilon_t}}
\newcommand{\ARMA}[1]{ARMA\ensuremath{(#1)}}
\newcommand{\ARIMA}[1]{ARIMA\ensuremath{(#1)}}
\newcommand{\AR}[1]{AR\ensuremath{(#1)}}
\newcommand{\MA}[1]{MA\ensuremath{(#1)}}
\newcommand{\sdf}{spectral density function}
\newcommand{\pdf}{probability density function}
\newcommand{\FI}[1]{FI\ensuremath{(#1)}}
\newcommand{\FID}[1]{\FAR{0,#1,0}}
\newcommand{\FAR}[1]{ARFIMA\ensuremath{(#1)}}
\newcommand{\dint}{\ensuremath{(-\frac{1}{2},\frac{1}{2})}}
\newcommand{\Sstandard}[2]{\ensuremath{\mathcal{S}_{#1,#2}}}
\newtheorem{Theorem}{Theorem}
\newtheorem{Corollary}{Corollary}
\newtheorem{Lemma}{Lemma}

\graphicspath{{../thesis/Figures/}}

\title{\vspace{-1cm} A brief history of long memory: Hurst, Mandelbrot and the road to ARFIMA}
\author{
  Timothy Graves\thanks{Arup, London, UK}
  \and
  Robert B.~Gramacy\thanks{ The University of Chicago Booth School of Business; 5807 S. Woodlawn Avenue, Chicago, IL 60637}
  \and
  Nicholas W.~Watkins\thanks{Corresponding author: Centre for
  Fusion Space and Astrophysics, University of Warwick, Coventry, UK;   Department of Engineering and Innovation, Faculty of Mathematics Computing and Technology, Open University, Milton Keynes, UK; Centre for the Analysis of Time Series,
  London School of Economics and Political Science, London, UK; and Max Planck Institute for the Physics of Complex
  Systems, Dresden, Germany ;
  {\tt NickWatkins@mykolab.com}}\\
    \and
  Christian L. E. Franzke\thanks{Meteorological Institute and Centre for Earth System Research and Sustainability (CEN), University of Hamburg, Germany}}

\date{}

\maketitle

\vspace{-0.5cm}

\begin{abstract}
Long memory plays an important role in many fields by determining the behaviour and predictability of systems; for instance, climate, hydrology, finance, networks and DNA sequencing. In particular, it is important to test if a process is exhibiting long memory since that impacts the accuracy and confidence with which one may predict future events on the basis of a small amount of historical data. A major force in the development and study of long memory was the late Benoit B. Mandelbrot. Here we discuss the original motivation of the development of long memory and Mandelbrot's influence on this fascinating field. We will also elucidate the sometimes
contrasting approaches to long memory in different scientific communities.\\

\end{abstract}

\doublespacing

\section{Introduction}
\label{sec:intro}

In many fields, there is strong evidence that a phenomenon called ``long memory" plays a significant role, with implications for forecast skill, low frequency variations, and trends. In a stationary time series
the term ``long memory"---sometimes ``long range dependence'' (LRD) or "long term persistence"---implies that there is non-negligible dependence between the present and {\em all} points in the past. 
To dispense quickly with some technicalities, we clarify here that our presentation
follows the usual convention in statistics \citep{Beran_1994,Beran_2013} and define a stationary finite variance
process to have long memory when its autocorrelation function (ACF) diverges: $\sum_{k=-\infty}^\infty \rho(k) = \infty$. This is equivalent to its power spectrum having a pole at zero frequency \citep{Beran_1994,Beran_2013}. In practice, this means the ACF and the power spectrum both follow a power-law, because the underlying process does not have any characteristic decay timescale. This is in striking contrast to many standard (stationary) stochastic processes where the effect of each data point decays so fast that it rapidly becomes indistinguishable from noise. 
The study of long memory processes is important because they exhibit nonintuitive properties where many familiar mathematical results fail to hold, and because of the numerous datasets \citep{Beran_1994,Beran_2013} where evidence for long memory has been found. In this paper we will give a historical account of three key aspects of long memory:
1) The environmetric observations in the 1950s which first sparked interest: the anomalous growth of range in hydrological time series, later known as the ``Hurst" phenomenon;
2) After more than  a decade of controversy, the introduction by Mandelbrot of the first stationary model-fractional Gaussian noise (FGN)-which could explain the Hurst phenomenon\footnote{This was in itself controversial because it explicitly  exhibited LRD (which he dubbed ``the Joseph effect")}; and 
3)The incorporation of LRD, via a fractional differencing parameter $d$, into the more traditional \ARMA{p,q} models, through Hosking and Granger's \FAR{p,d,q} model.

The development of the concept of long memory, both as a physical notion and a
formal mathematical construction, should be of significant interest in the light
of controversial  application areas like the study of bubbles and business cycles  in
financial markets \citep{Sornette_2004}, and the quantification of climate trends \citep{Franzke:2012}. Yet few
articles about long memory cover the history in much  detail. Instead most
introduce the concept with passing reference to its historical
significance; even books on LRD tend to have only a brief
history.\footnote{Notable exceptions include \citet{Montanari_2003}, 
the semi-autobiographical \citet{Mandelbrot_2008}, and his posthumous autobiography \citep{Mandelbrot_2013}, as well as the reminiscence of his former student Murad \cite{Taqqu_2013}} This lack of historical context is important not just because a knowledge of the history is intrinsically rewarding, but also because understanding the conceptual development of a research field  can help to avoid pitfalls in future. Here we attempt to bridge the gap in a way that is both entertaining and accessible to a wide statistical and scientific audience.
We assume no mathematical details beyond those given in
an ordinary time series textbook \citep[e.g.,][]{Brockwell_1991},
and any additional notation and concepts will be kept to a minimum.

The key questions that we seek to answer are ``Who first considered long
memory processes in time series analysis, and why?'' and ``How did these early
studies begin to  evolve into the modern-day subject?''\footnote{For specificity, we
clarify here that our interpretation of
`modern-day subject' comprises of the definitions of long memory given 
above, including footnoted alternatives, 
together with the ARFIMA$(p,d,q)$ processes defined through the backshift 
operator $\mathcal{B}$ as 
$\Phi(\mathcal{B}(1-\mathcal{B})^d X_t = \Theta(\mathcal{B}) \varepsilon_t$
where $\Phi$ and $\Theta$ are autoregressive and moving average polynomials,
and $\varepsilon_t$ is white noise.  For more details, see any modern
time series text \citep[e.g.,][]{Brockwell_1991}.} As we
shall see, this evolution took less than three decades across the middle of
the twentieth century. During this period there was significant debate about
the mathematical, physical, and philosophical interpretations of long memory.
It is both the evolution of this concept, and the accompanying debate (from
which we shall often directly quote), in which we are mostly interested. The kind of memory that concerns us here was a conceptually new idea in science, and rather different, for example, from that embodied in the laws of motion developed by Newton and Kepler. Rather than Markov processes where the current state of a system now is enough to determine its immediate future, the fractional Gaussian noise model requires information about the complete past history of the system.

As will become evident, the late Beno\^{\i}t B. Mandelbrot was a key
figure in the development of long memory. Nowadays most famous for coining the
term and concept `fractal', Mandelbrot's   output 
  crossed a wide variety of
subjects from hydrology to economics as well as pure and applied mathematics.
During the 1960s he worked on the theory
of stochastic processes exhibiting heavy tails and long memory, and was the
first to distinguish between these effects.  Because of the diversity of the communities in which he made contributions, it sometimes seems that Mandelbrot's role in statistical modelling is perhaps underappreciated (in contrast, say, to within the physics and geoscience communities \citep{AharonyFeder1990,Turcotte1997}). It certainly seemed this way to him:
\begin{quotation} \singlespacing
 Of those three [i.e. economics, engineering, mathematics], nothing beats my impact on finance and mathematics. Physics - which I fear was least affected - rewarded my work most handsomely.  \citep{Mandelbrot_2013}
\end{quotation} 
 A significant portion of this paper is devoted to his work. We do not, however, intend to convey in any sense his `ownership' of
the LRD concept, and indeed much of the modern progress concerning long memory in statistics has adopted an approach (ARFIMA) that he did not   agree with. 

Mandelbrot's motivation in developing an interest in long memory processes
stemmed from an intriguing study in hydrology by Harold \citet{Hurst_1951}.
Before we proceed to discuss this important work it is necessary to give a
brief history of hydrological modelling,  Hurst's contributions, and the reactions to him
from other authors in that
area in Section \ref{sec:hydro}.  Then we discuss Mandelbrot's initial
musings, his later refinements, and the reactions from the hydrological community in Section 
\ref{section: Mandelbrot's model}. 
In Section \ref{section: fractionally differenced models} we discuss the development in the 1980s of fractionally differenced models culminating from this
sequence of thought.  Section \ref{sec:conclude} offers our conclusions.

\section{Hurst, and a brief history of hydrology models}
\label{sec:hydro}
Water is essential for society to flourish since it is required for drinking,
washing, irrigation and for fuelling industry. For thousands of years going
back to the dawn of settled agricultural communities, humans have sought 
methods to regulate the natural flow of water. They tried to
control nature's randomness by building reservoirs to store water in times of
plenty, so that lean times are survivable. The combined factors of
the nineteenth century Industrial Revolution, such as fast urban population
growth, the requirement of mass agricultural production, and increased energy
requirements, led to a need to build large scale reservoirs formed by the
damming of river valleys. When determining the capacity of the reservoir, or
equivalently the height of the required dam, the natural solution is the
`ideal dam': 
\begin{quotation} \singlespacing
[An `ideal dam' for a given time period is such
that] (a) the outflow is uniform, (b) the reservoir ends the period as full as
it began, (c) the dam never overflows, and (d) the capacity is the smallest
compatible with (a), (b) and (c). \citep{Mandelbrot_1969a}\footnote{The
concept of the ideal dam obviously existed long before Mandelbrot, however the quotation is a
succinct mathematical definition. Naturally, this neat mathematical
description ignores complications such as margins of error, losses due to
evaporation etc.\ but the principle is clear. Actually, as \citet
{Hurst_1951} himself pointed out: ``increased losses due to storage are
disregarded because, unless they are small, the site is not suitable for
over-year storage''.} 
\end{quotation}

From a civil engineer's perspective, given the parameters of demand (i.e.\
required outflow) and time horizon, how should one determine the optimal
height of the dam? To answer this question we clearly need an input, i.e.\
river flows. It is not hard to imagine that for a given set of inputs it
would, in principle, be possible to mathematically solve this problem. A
compelling solution was first considered by
\citet{Rippl_1883} ``whose publication can \ldots be identified with the
beginning of a rigorous theory of storage reservoirs'' \citep
{Klemes_1987}.

Despite solving the problem, \citeauthor{Rippl_1883}'s method was clearly
compromised by its requirement to know, or at least assume, the future variability of the river flows. A
common method was to use the observed history at the site as a proxy; however
records were rarely as long as the desired time horizon. Clearly a stochastic
approach was required, involving a simulation of the future using a
stochastic process known to have similar statistical properties to the
observed past. This crucial breakthrough, heralding the birth of stochastic
hydrology, was made by \citet{Hazen_1914} who used the simplest possible model; an \iid{} Gaussian process.

In practice, just one sample path would be of little use so, in principle,
many different sample paths could be generated, all of which could be analysed
using \citeauthor{Rippl_1883}'s method to produce a distribution of `ideal
heights'. This idea of generating repeated samples was pursued by
\citet{Sudler_1927}, however the stochastic approach to reservoir design was
not generally accepted in the West until the work of Soviet engineers was
discovered in the 1950s. The important works by \citet{Moran_1959} and
\citet{Lloyd_1967} are jointly considered to be the foundations of modern
reservoir design, and helped establish this approach as best practice.

\subsection{Hurst's paper}
\label{section: Hurst's paper}

Harold Edwin Hurst had spent a long career in Egypt (ultimately spanning
1906--68) eventually becoming Director-General of the Physical Department
where he was responsible for, amongst other things, the study of the
hydrological properties of the Nile basin.
For
thousands of years the Nile had helped sustain civilisations in an otherwise
barren desert, yet its regular floods and irregular flows were a severe
impediment to development. Early attempts at controlling the flow by damming
at Aswan were only partially successful. Hurst and his department were tasked
with devising a method of water control by taking an holistic view of the Nile
basin, from its sources in the African Great Lakes and Ethiopian plains, to
the grand delta on the Mediterranean.

In his studies of river flows, \citet{Hurst_1951} used a method similar to \citeauthor{Rippl_1883}'s in which he analysed a particular statistic of the cumulative flows of rivers over time called the `adjusted range', $R$. Let $\{X_k\}$ be a sequence of random
variables, not necessarily independent, with some non-degenerate distribution.
We define the $n$th partial sum $ Y_n =: X_1+ \cdots + X_n$. \citet{Feller_1951} then defines
the Adjusted Range, $R(n)$, as:
\begin{equation*}
R(n) = \max_{1\leq k\leq n}\left\{Y_k - \frac{k}{n}Y_n\right\} 
- \min_{1\leq k\leq n}\left\{Y_k - \frac{k}{n}Y_n\right\}.
\end{equation*}
Hurst referred to this as simply the `range' which is now more commonly used
for the simpler statistic $R^*(n) = \max_{1\leq k\leq n}\{Y_k\} - \min_{1\leq
k\leq n}\{Y_k\}$. Moreover he normalised the adjusted range by the sample
standard deviation to obtain what is now called the Rescaled Adjusted Range
statistic, denoted $R/S(n)$: 
\begin{equation*}
R/S(n) = \frac{\max_{1\leq k\leq n}\left\{Y_k - \frac{k}{n}Y_n\right\} - \min_{1\leq k\leq n}\left\{Y_k - \frac{k}{n}Y_n\right\}}{\sqrt{\frac{1}{n}\sum_{k=1}^{n}\left(X_k-\frac{1}{n}Y_n\right)^2}}.
\end{equation*}
Hurst's reasons for this normalisation are unclear but, as we shall see later,
proved remarkably fortuitous. The attraction of using $R/S$ is that, for a
given time period of say $n$ years, $R/S(n)$ is a proxy for the ideal dam
height over that time period.

\citet{Hurst_1951} 
then examined 690 different time series, covering 75 different geophysical
phenomena spanning such varied quantities as river levels, 
rainfall, temperature, atmospheric pressure, tree rings, mud sediment
thickness, and sunspots. He found that in each case, the statistic behaved as
$R/S(n) \propto n^k$ for some $k$. He estimated $k$ using a statistic he
called $K$, and found that $K$ was approximately normally distributed with
mean 0.72 and standard deviation 0.006. He 
actually acknowledged that ``$K$ does vary slightly with different phenomena'',
and that the range ($0.46 - 0.96$) was large for a Gaussian fit, however to a
first approximation it appeared that the mean value of $0.72$ might hold some
global significance.

At this point it is worth highlighting an aspect of Hurst's work which often
gets overlooked. As we shall see, the $R/S$ statistic has enjoyed great use
over the past fifty years. However the modern method of estimating the
exponent $k$ is not that originally used by Hurst. His estimate $K$ was
obtained by assuming a known constant of proportionality: specifically he
assumed the asymptotic  (i.e.\ for large $n$) law that $R/S(n) = (n/2)^k$. A
doubly logarithmic plot of values of $R/S(n)$ against $n/2$ should produce a
straight line, the slope of which is taken as $K$. By assuming a \emph{known}
constant of proportionality, Hurst was effectively performing a one parameter
log-regression to obtain his estimate of $k$.

His reason for choosing this approach was that it implies $R/S(2)=1$
exactly\footnote{it actually equals $1/\sqrt{2}$ but Hurst was calculating
standard deviations by dividing by $n$ rather than $n-1$}, and consequently
this `computable value' could be used in the estimation procedure. This
methodology would nowadays be correctly regarded as highly dubious because it
involves fitting an asymptotic (large $n$) relationship while making use of an assumed small fixed value for the $n=1$ point.
This logical flaw was immediately remarked upon in the same journal issue by \cite{Te_Chow_1951}. As we
will see, Mandelbrot later introduced the now-standard method of estimation by
dropping this fixed point and performing a \emph{two}-parameter log-regression
to obtain the slope. Hurst's original method was forgotten and most
authors are unaware that it was not the same as the modern method; indeed many
cite Hurst's result of $0.72$ unaware that is was obtained
using an inappropriate analysis.

Notwithstanding these shortcomings, \citeauthor{Hurst_1951}'s
key result that estimates of $k$ were about 0.72 would likely
not have been either noteworthy or controversial in itself had he not shown
that, using contemporary stochastic models, this behaviour could not be
explained.

In the early 1950s, stochastic modelling of river flows was
immature and so the only model that Hurst could consider was the \iid{}
Gaussian model of \citet{Hazen_1914} and \citet{Sudler_1927}. Rigorously deriving the
distribution of the range under this model was beyond Hurst's mathematical skills, but by
considering the asymptotics of a coin tossing game and appealing to the central
limit theorem, he did produce an extraordinarily good heuristic solution. His
work showed that, under the independent Gaussian assumption, the exponent $k$
should equal 0.5. In other words Hurst had shown that contemporary
hydrological models fundamentally did not agree with empirical evidence. This
discrepancy between the theory and practice became known as the `Hurst
phenomenon'.\footnote{It is worth clarifying a potential ambiguity here: since
the phrase was coined, the `Hurst Phenomenon' has been attributed to various
aspects of time series and/or stochastic processes. For clarity, we will use
the term to mean ``the statistic $R/S(n)$ empirically grows faster than
$n^{1/2}$''.}

Hurst's observation sparked a series of investigations that ultimately led to
the formal development of long memory. Hurst himself offered no specific
explanation for the effect although he clearly suspected the root of cause
might lie in the independence assumption: 
\begin{quotation} \singlespacing
Although in random
events groups of high or low values do occur, their tendency to occur in
natural events is greater. ... There is no obvious periodicity, but there are
long stretches when the floods are generally high, and others when they are
generally low. These stretches occur without any regularity either in their
time of occurrence or duration \citep[\S6]{Hurst_1951}.
\end{quotation}
Despite several follow-up publications
\citep{Hurst_1956,Hurst_1956a,Hurst_1965},  Hurst
himself played no direct part in the further development of long memory. The specific
purpose of his research was to design a system to control the Nile with a
series of small dams and reservoirs. These plans were later turned into the
Aswan High Dam with Hurst still acting as scientific consultant into his
eighties \citep{Mandelbrot_2008}.

\subsection{Reactions to the Hurst phenomenon}
\label{sec:hreact}
Hurst's finding took the hydrological community by surprise, not only because
of the intrinsic puzzle, but because of its potential importance. As
previously mentioned, the $R/S(n)$ statistic is a proxy for the ideal dam
height over $n$ years. If Hurst's finding was to be believed, and $R/S(n)$ increased  \emph{faster} than $n^{1/2}$, there would be potentially
major implications for dam design. In other words, dams designed for long time
horizons might be too low, with floods as one potential consequence.

Although the debate over Hurst's findings, which subsequently evolved into the
debate about long memory, was initially largely confined to the hydrological
community, fortuitously it also passed into more mainstream mathematical
literature --- a fact which undoubtedly helped to lend it credence in later
years. Despite Hurst's non-rigorous approach, and an unclear mathematical
appeal for what was essentially a niche subject, the eminent probabilist William
\citet{Feller_1951} contributed greatly by publishing a short paper. By
appealing to Brownian motion theory, he proved that Hurst was
correct; for sequences of standardised \iid{} random variables with finite
variance, the \emph{asymptotic} distribution of the adjusted range, $R(n)$,
should obey the $n^{1/2}$ law:
$\mathbbm{E}[R(n)] \sim \left(\frac{\pi}{2}\right)^{1/2}n^{1/2}$.
It should be emphasised that \citeauthor{Feller_1951} was studying the
distribution of the adjusted range, $R(n)$, not the \emph{rescaled} adjusted
range $R/S(n)$. The importance of dividing by the standard deviation was not
appreciated until Mandelbrot, however \citeauthor{Feller_1951}'s results would
later be shown to hold for this statistic as well.

By proving and expanding (since the Gaussianity assumption could be weakened)
Hurst's result, \citeauthor{Feller_1951} succeeded in both confirming that
there was a phenomenon of interest, and also that it should interest
mathematicians as well as hydrologists. Over the course of the 1950s more
precise results were obtained although attention was unfortunately deflected to
consideration of the simple range
\citep[e.g.][]{Anis_1953}
as opposed to $R/S$. The exact
distribution of $R(n)$ was found to be, in general, intractable; a notable
exception being that for the simplest \iid{} Gaussian case, where  \citep{Solari_1957}
\begin{equation*}
\mathbbm{E}[R(n)] = \left(\frac{\pi}{2}\right)^{1/2} 
\left(\frac{1}{\pi}\sum_{k=1}^{n-1}\frac{1}{\sqrt{k(n-k)}}\right) n^{1/2}.
\end{equation*} 
Having conclusively shown that Hurst's findings were indeed worthy of
investigation, several different possible explanations of the eponymous phenomenon
were put forward. It was assumed that the effect was caused by (at least) one
of the following properties of the process: a) an `unusual' marginal
distribution, b) non-stationarity, c) transience (i.e\ pre-asymptotic
behaviour), or d) short-term auto-correlation effects.

For Hurst's original data, the first of these proposed solutions was not
relevant because much of his data were clearly Gaussian. \cite{Moran_1964}
claimed that the effect could be explained by using a sequence of \iid{}
random variables with a particular moment condition on the distribution.
Although this case had been shown by \citet{Feller_1951} to still
\emph{asymptotically} produce the $n^{1/2}$ law, Moran showed that in such
cases the transient (or the \emph{pre}-asymptotic) phase exhibiting the Hurst
phenomenon could be extended arbitrarily.\footnote{Moran used a Gamma
distribution, although to achieve the effect the distribution had to be
heavily skewed, thus ruling it out as a practical explanation for Hurst's
effect.} Furthermore, Moran pointed out that if the finite variance assumption
was dropped altogether, and instead a symmetric $\alpha$-stable distribution
was assumed, the Hurst phenomenon could apparently be explained:
$\mathbbm{E}[R(n)] \sim \ell n^{1/\alpha}$, for $1 < \alpha \leq 2$.
and some known (computable) $\ell$. 
However, as Mandelbrot later showed, the division by the standard deviation is
indeed crucial. In other words, whilst \citeauthor{Moran_1964}'s arguments
were correct, they were irrelevant because the object of real interest was the
\emph{rescaled} adjusted range. Several Monte Carlo studies, notably those by
\citet{Mandelbrot_1969c}, confirmed that for \iid{} random variables,
regardless of marginal distribution, $R/S(n)$ asymptotically follows the
$n^{1/2}$ law. That \citet{Mandelbrot_1975} and
\citet{Mandelbrot_1979} were finally able to prove this result remains
of the principal reasons why the $R/S$ statistic has remained popular to the
present day.

The second potential explanation of the Hurst phenomenon, non-stationarity, is
harder to discount and is more of a philosophical (and physical) question than a mathematical
one.  
\begin{quotation} \singlespacing
Is it meaningful to talk of a time-invariant mean over thousands of years? If
long enough realizations of such time series were available would they in fact
be stationary? \citep[\S3.2]{OConnell_1971}
\end{quotation} 
Once we assume the possibility of non-stationarity, it is not hard to imagine
that this could lead to an explanation of the phenomenon. Indeed
\citet{Hurst_1957} himself suggested that non-stationarity might be an
explanation; however his heuristics involving a pack of playing cards were far
from being mathematically formalisable. \citet{Klemes_1974} and
\citet{Potter_1976} later provided more evidence, however the first rigorous
viable mathematical model was that by \citet{Bhattacharya_1983}, in which the
authors showed that a short-memory process perturbed by a non-linear monotonic
trend can be made to exhibit the Hurst phenomenon.\footnote{Their study shows
why it is crucial to distinguish between the `Hurst phenomenon' and `long
memory'. The process described by \citeauthor{Bhattacharya_1983} does not have
long memory yet it exhibits the Hurst phenomenon (recall our specific
definition of this term).}

\label{quote: Klemes}
In his influential paper, \citet{Klemes_1974} not only showed that the
Hurst phenomenon could be explained by non-stationarity, but argued that
assuming stationarity may be mis-founded:
\begin{quotation} \singlespacing
The question of whether natural processes are stationary or not is likely a
philosophical one. \ldots there is probably not a single historic time series
of which mathematics can tell with certainty whether it is stationary or not
\ldots Traditionally, it has been assumed that, in general, the geophysical,
biological, economical, and other natural processes are nonstationary but
within relatively short time spans can be well approximated by stationary
models. \citep{Klemes_1974}
\end{quotation} 
As an example,
\citeauthor{Klemes_1974} suggested that a major earthquake might drastically
affect a river basin so much as to induce a regime-change (i.e.\ an element of
non-stationarirty). However on a larger (spatial and temporal) scale, the
earthquake and its local deformation of the Earth may be seen as part of an
overall stationary `Earth model'. Thus choosing between the two forms is, to
some extent, a matter of personal belief. As we will see in the next section, Mandelbrot did in fact consider (and publish) other   models with a particular type of nonstationary switching himself, even while formulating his stationary FGN model, but unfortunately \citeauthor{Klemes_1974} was unaware of that work, about which a more fruitful discussion might perhaps have occurred. 

If we discount this explanation and assume stationarity, we must turn
to the third and fourth possible explanations, namely transience (i.e.\
pre-asymptotic behaviour) and/or the lack of independence. These two effects
are related: short-term auto-correlation effects are likely to introduce
significant pre-asymptotic behaviours. As mentioned earlier, Hurst himself
suggested some kind of serial dependence might explain the effect, and
\citeauthor{Feller_1951} suggested: 
\begin{quotation} \singlespacing
It is conceivable that
the [Hurst] phenomenon can be explained probabilistically, starting from the
assumption that the variables $\{X_k\}$ are not independent \ldots
Mathematically this would require treating the variables $\{X_k\}$ as a Markov
process. \citep{Feller_1951} 
\end{quotation}

Soon however, \citet{Barnard_1956} claimed to have shown that Markovian models
still led to the $n^{1/2}$ law and it would be shown later
\citep{Mandelbrot_1968b,Mandelbrot_1975} that any then-known form of
auto-correlation must asymptotically lead to the same result. The required
condition on the auto-correlation function turned out to be that it is
summable, whereby 
for \iid{} random variables with \acf{} $\rho(\cdot)$ \citep{Siddiqui_1976}:
\begin{equation}
\label{eqn: asymptotic distribution of RAR} 
\mathbbm{E}[R/S(n)] \sim 
\left(\frac{\pi}{2}\right)^{1/2}\left(\sum_{k=-\infty}^{\infty}\rho(k)\right)^{1/2} 
n^{1/2}.
\end{equation}
Even before this was formally proved, it was generally known that some
complicated auto-correlation structure would be necessary to explain the Hurst
phenomenon: 
\begin{quotation} \singlespacing
It has been suggested that serial correlation or
dependence [could cause the Hurst phenomenon]. This, however, cannot be true
unless the serial dependence is of a very peculiar kind, for with all
plausible models of serial dependence the series of values is always
approximated by a [Brownian motion] when the time-scale is sufficiently large.
A more plausible theory is that the experimental series used by
\citeauthor{Hurst_1951} are, as a result of serial correlation, not long
enough for the asymptotic formula to become valid.  \citep
{Moran_1959}
\end{quotation} 
Thus Moran was arguing that, since no `reasonable' auto-correlation structure
could account for the Hurst phenomenon, it should be assumed that the observed
effect was caused by pre-asymptotic behaviour, the extent of which was
influenced by some form of local dependence. In other words, he was suggesting
that a short-memory process could account for the Hurst phenomenon over
observed time scales.

This issue has both a practical and philosophical importance. It would later
be argued by some that, regardless of the `true' model, any process that could
exhibit the Hurst phenomenon over the observed (or required) time scales would
suffice for practical purposes. Using such processes requires a choice. One
might accept the Hurst phenomenon as genuine and acknowledge that, although
theoretically incorrect, such a model is \emph{good enough} for the desired
purpose. Alternatively, one might reject the Hurst phenomenon as simply a
pre-asymptotic transient effect, and therefore any model which replicates the
effect over observed ranges of $n$ is potentially valid.
Mandelbrot, for one, was
highly critical of those who followed the latter approach:
\begin{quotation} \singlespacing
So far, such a convergence [to the $n^{1/2}$ law] has never been observed in
hydrology. Thus, those who consider Hurst's effect to be transient implicitly
attach an undeserved importance to the value of [the sample size] \ldots These
scholars condemn themselves to never witness the full asymptotic development
of the models they postulate. \citep{Mandelbrot_1968b}
\end{quotation}
Despite this, the concept of short-memory-induced transience was explored both
before and after Mandelbrot's work. \citet{Matalas_1967} performed a rigorous
Monte Carlo analysis of the \AR{1} model and demonstrated that for medium $n$
and heavy lag-one serial correlation, the Hurst phenomenon could be induced
(albeit \citeauthor{Matalas_1967} were actually using Hurst's original
erroneous $K$ estimate). \citet{Fiering_1967} succeeding in building a more
sophisticated model; however he found he needed to use an \AR{20} process to
induce the effect --- an unrealistically large number of lags to be useful for
modelling.

To summarise, by the early 1960s, more than a decade on from Hurst's original
discoveries, no satisfactory explanation for the Hurst
phenomenon had yet been found. To quote \citep{Klemes_1974} again:
\begin{quotation} \singlespacing
Ever since \citeauthor{Hurst_1951} published his famous plots for some
geophysical time series \ldots the by now classical Hurst phenomenon has
continued to haunt statisticians and hydrologists. To some it has become a
puzzle to be explained, to others a feature to be reproduced by their models,
and to others still, a ghost to be conjured away.  
\end{quotation} 
It was at this point that Beno\^{\i}t Mandelbrot heard of the phenomenon.

\section{Mandelbrot's fractional models}
\label{section: Mandelbrot's model}
\setcounter{footnote}{0}
In the early 1960s, Mandelbrot had worked intensively on the burgeoning
subject of mathematical finance which is concerned with modelling economic
indices such as share prices. Central to this subject was the `Random Walk
Hypothesis' which provided for Brownian motion models. This was first
implicitly proposed in the seminal (yet long undiscovered) doctoral thesis by
\cite{Bachelier_1900}. The detailed development of this topic is also
interesting but beyond the scope of this paper. It suffices to say here
that, although \citeauthor{Bachelier_1900}'s model was recognised as an
adequate working model which seemed to conform to both intuition and the data,
it could also benefit from refinements. Various modifications were proposed
but one common feature they all shared was the underlying Gaussian assumption.

In a ground-breaking paper, \citet{Mandelbrot_1963} proposed dropping the
Gaussianity assumption and instead assuming a heavy tailed distribution,
specifically the symmetric $\alpha$-stable distribution
\citep[e.g.,][\S1.1]{Samorodnitsky_1994}. In short, this notion was highly
controversial; for example see \citet{Cootner_1964b}. But the paper was
significant for two reasons. Firstly it helped to give credibility to the
growing study of heavy tailed distributions and stochastic processes. Secondly
it was indicative of Mandelbrot's fascination with mathematical scaling. The
$\alpha$-stable distributions have the attractive property that an
appropriately re-weighted sum of such random variables is itself an
$\alpha$-stable random variable. This passion for scaling would remain with
Mandelbrot throughout his life, and is epitomised by his famous fractal geometry.

Returning to Hurst's results, Mandelbrot's familiarity with scaling helped him
immediately recognise the Hurst phenomenon as symptomatic of this, and, 
as he later recounted \citep{Mandelbrot_2002a,Mandelbrot_2008}, he 
assumed that it could be explained by heavy tailed processes. He was therefore
surprised when he realised that, not only were Hurst's data essentially
Gaussian, but as discussed previously, the \emph{rescaled} adjusted range is
not sensitive to the marginal distribution. 
Instead, he realised that a new approach would be required. In keeping with
the idea of scaling, he introduced the term
`self-similar',\footnote{Mandelbrot later regretted the term `self-similar'
and came to prefer `self-affine', because scaling in time and space were not
necessarily the same, but the revised terminology never caught on to the same extent.} 
and formally introduced the concept in its modern
form: Let $Y(t)$ be a continuous-time stochastic process. Then $Y(t)$ is said to be
self-similar, with self-similarity parameter $J$, if for all positive $c$, $Y(ct)\stackrel{d}{=}c^J Y(t)$.
Using this concept, \citet{Mandelbrot_1965}, laid the foundations for the
processes which would initially become the paradigmatic models  in the field of long memory, the self-similar
fractional Brownian motion (FBM) model and its increments, the long range dependent fractional Gaussian noise (FGN) model.

At this point it is necessary to informally describe FBM. It is a
continuous-time Gaussian process, and is a generalisation of ordinary Brownian
motion, with an additional parameter $h$.\footnote{We remark that the naming
and notation of this parameter has been a source of immense confusion over the
past half-century, with various misleading expressions such as the `Hurst
parameter', the `Hurst exponent', the `Hurst coefficient', the
`self-similarity parameter' and the `long memory parameter'. Moreover, the
more traditional notation of an upper-case $H$ does not help since it disobeys
the convention of using separate cases for constants (parameters) and random
variables (statistics). For clarity in what follows we will
we will reserve the notation $h$ simply to denote the `fractional
Brownian motion parameter'.} This  parameter can range between zero and one
(non-inclusive to avoid pathologies) with different values providing
qualitatively different types of behaviour. The case $h=1/2$ corresponds to
standard Brownian motion.

Fractional Brownian motion can be thought of in several different and
equivalent ways, for example as a fractional derivative of standard Brownian
motion, or as stochastic integral. These details need not concern us here; the
most important fact is that FBM is exactly self-similar, which means that a
`slowed-down' version of a process will, after a suitable spatial re-scaling,
look \emph{statistically} identical to the original, i.e.\ they will have the
same finite dimensional distributions. In this sense FBM, like standard
Brownian motion (which of course is just a special case of FBM), has no
characteristic time-scale, or `tick'.

In practical applications it is necessary to use a modification of FBM because
it is (like standard Brownian motion) a continuous time process and
non-stationary. Thus the increments of FBM are considered; these form a
discrete process which can be studied using conventional time series analysis
tools. These increments, called fractional Gaussian noise (FGN), can be
considered to be the discrete approximation to the `derivative' of FBM. Note
that in the case of $h=1/2$, FGN is simply the increments of standard Brownian
motion, i.e.\ white noise. 
\citet[][corollary 3.6]{Mandelbrot_1968a}
showed that this process is stationary, but most
importantly (for its relevance here), it exhibits the
Hurst phenomenon: for some $c>0$, $ R/S(n) \sim cn^h$.
This result was immensely significant; it was the first time since Hurst had
first identified the phenomenon, that anyone had been able to exhibit a
stationary, Gaussian stochastic process capable of reproducing the effect. The
mystery had been partially solved; there \emph{was} such a process, and for
over a decade it remained the only model known to be able to fully explain the
Hurst phenomenon.

\citet{Mandelbrot_1965} then proceeded to show
that such a process must have a \sdf{} that blows up at the origin. By
proposing such a model, he realised he was attempting to explain with one
parameter $h$ both low- and high-frequency effects, i.e.\ he was ``\ldots
postulating the same mechanism for the slow variations of climate and for the
rapid variations of precipitation''. He also recognised that the
auto-correlation function of the increments would decay slower than
exponentially, and (for $1/2<h<1$) would not be summable. This correlation
structure, which is now often taken to be the definition of long memory
itself, horrified some. Concurrently, the simplicity of FGN, possessing only
one parameter $h$, concerned others. We shall consider these issues in depth
later, but the key point was that although Mandelbrot had `conquered' the
problem, to many it was somewhat of a Pyrrhic victory \citep{Klemes_1974}.

\subsection{Initial studies of Mandelbrot's model}
\label{sec:initman}

Mandelbrot immediately attempted to expand on the subject although his papers
took time to get accepted. He ultimately published a series of five papers in
1968--9 through collaborations with the mathematician John Van Ness and the hydrologist James Wallis. Taken as a whole, these papers offered a
comprehensive study of long memory and fractional Brownian motion. They helped
publicise the subject within the scientific community and started the debates
about the existence of long memory and the practicality of FGN, which have
continued until the present day.

The first of these papers \citep{Mandelbrot_1968a} formally introduced FBM and
FGN and derived many of their properties and representations. The aim of this
paper was simply to introduce these processes and to demonstrate that they
could provide an explanation for the Hurst phenomenon; this was succinctly
stated:
\begin{quotation} \singlespacing
We believe FBM's do provide useful models for a host of natural time series
and wish therefore to present their curious properties to scientists,
engineers and statisticians.
\end{quotation} 
\citeauthor{Mandelbrot_1968a} argued that all processes thus far considered
have ``the property that sufficiently distant samples of these functions are
independent, or nearly so'', yet in contrast, they pointed out that FGN has
the property ``that the `span of interdependence' between their increments can
be said to be infinite''. This was a qualitative statement of the difference
between short and long memory and soon led to the formal definition of long
memory. As motivation for their work, they cited various examples of observed
time series which appeared to possess this property: in economics
\citep{Adelman_1965,Granger_1966}, `$1/f$ noises' in the fluctuations of
solids \citep{Mandelbrot_1967c}, and hydrology \citep{Hurst_1951}.

Intriguingly, and undoubtedly linked to Mandelbrot's original interest in heavy-tailed processes, \citet[\S3.2]{Mandelbrot_1968a} noted:
\begin{quotation} \singlespacing
If the requirement of continuity is abandoned, many other
interesting self-similar processes suggest themselves. One may for example
replace [the Brownian motion] by a non-Gaussian process whose increments are
[$\alpha$-] stable \ldots Such increments necessarily have an infinite
variance. `Fractional L\'{e}vy-stable random functions' have moreover an
infinite span of interdependence. 
\end{quotation}

In other words the authors postulated a heavy-tailed, long memory process. It
would be over a decade before such processes were properly considered
due to difficulties arising from the lack of formal
correlation structure in the presence of infinite variance.\footnote{However a preliminary demonstration of the robustness of R/S as a measure of LRD was given in \cite{Mandelbrot_1969c}, using a heavy tailed modification of fBm which they dubbed ``fractional hyperbolic motion".}

One key point which is often overlooked is that
\citeauthor{Mandelbrot_1968a} did not claim that FGN is necessary to explain
the Hurst phenomenon: ``\ldots we selected FBM so as to be able to derive the
results of practical interest with a minimum of mathematical difficulty''.
Often Mandelbrot was incorrectly portrayed as insisting that his, and
\emph{only} his, model solved the problem. Indeed Mandelbrot himself took an interest in alternative models, although as we will later see, he
essentially rejected   Granger and Hosking's ARFIMA which  was to become the standard replacement of FGN in statistics and econometrics literatures.

Furthermore, neither did the authors claim
that they were the first to discover FBM. They acknowledged that others
\citep[e.g.][]{Kolmogorov_1940} had implicitly studied it; however
\citeauthor{Mandelbrot_1968a} were undoubtedly the first to attempt to use it
in a practical way.

Having `solved' Hurst's riddle with his stationary fractional Gaussian model, 
Mandelbrot determined to get  FGN and FBM
studied and accepted, in particular by the community which had most interest
in the phenomenon, hydrology. Therefore his remaining four important papers
were published in the leading hydrological journal \emph{Water Resources
Research}. These papers represented a comprehensive study of FBM in an
applied setting, and were bold; they called for little short of a revolution
in stochastic modelling: 
\begin{quotation} \singlespacing
... current models of statistical hydrology cannot account for either [Noah or
Joseph] effect and must therefore be superseded. As a replacement, the
`self-similar' models that we propose appear very promising
\citep{Mandelbrot_1968b}
\end{quotation}
As its title suggests, \citet{Mandelbrot_1968b} introduced the colourful terms
`Noah Effect' and `Joseph Effect' for heavy tails and long memory
respectively; both labels referencing key events of the Biblical characters'
lives. Ironically,  the river level data were in fact close enough to Gaussian
to dispense with the `Noah Effect' so the actual content of the paper was
largely concerned with the `Joseph Effect', but rainfall itself provides a rich source of heavy tailed, ``Noah" datasets. However 
Mandelbrot preferred treating these two effects together as different forms of
scaling; spatial in the former and temporal in the latter.

\citet{Mandelbrot_1968b} defined the `Brownian domain of attraction' (BDoA)
and showed that such BDoA cannot account for either effect and should
therefore be discarded. The BDoA was (rather loosely) defined as the set of
discrete-time stochastic processes which obey three conditions; namely, the
Law of Large Numbers, the Central Limit Theorem, and asymptotic independence
of past and future partial sums. Alternatively the BDoA is the set of
processes which are either asymptotically Brownian, or can be
well-approximated by Brownian motion. A process in the BDoA is in some sense
`nice', i.e.\ it is Gaussian or Gaussian-like and has short memory, and was
given the term `smooth'. Processes outside of the BDoA were labelled
`erratic'\footnote{Mandelbrot later preferred the terms `mild' to ``nice", and subdivided ``erratic" into heavy tailed `wild' and strongly dependent `slow". We  stick with his original terminology.}.
This `erratic' behaviour could be caused by one, or both, of the Joseph and
Noah effects. \citeauthor{Mandelbrot_1968b} showed that processes lying within
the BDoA will, after an initial transient behaviour, obey the $n^{1/2}$ law.
They rejected, on philosophical grounds, the idea that the Hurst phenomenon
might be caused by transient effects.

Mandelbrot proceeded to provide more evidence in support of his model.
\citet{Mandelbrot_1969a} included several sample graphs of simulated
realisations of FGN with varying $h$. The explicit aim was to ``encourage
comparison of [the] artificial series with the natural record with which the
reader is concerned''. These simulations were performed by using one of two
methods developed by the authors which were different types of truncated
approximations.\footnote{As documented by \citet{Mandelbrot_1971} it was soon
found that one of these approximations was far from adequate because it failed
to accurately reproduce the desired effects. The algorithms were also slow to
implement; a significant practical problem when computer time was expensive
and processing power limited. \citet{Mandelbrot_1971} therefore introduced a
more efficient algorithm. Later an \emph{exact} algorithm would be created by
\citet{Hipel_1978b,Hipel_1978c} which forms the basis for modern algorithms%
\citep{Davies_1987,Beran_1994} which use the Fast Fourier Transform
.} \citeauthor{Mandelbrot_1969a} wanted to subject these simulations to $R/S$
analysis but they recognised the previously mentioned logical flaw in Hurst's
approach. They therefore developed a systematic two-parameter log-regression
to obtain an estimate of $h$ which we will denote $H$. This approach to $R/S$
analysis has since become the standard method.

The simulated sample paths were subjected to both $R/S$ and spectral analysis,
and for both cases it was found that the simulated paths largely agreed with
the theory, i.e.\ the sample paths seemed sufficiently good representations of
the theoretical processes. For the $R/S$ analysis, it was found, as expected,
that there existed three distinct regions: transient behaviour, `Hurst'
behaviour, and asymptotic `1/2' behaviour. This last region was caused
entirely because the simulations were essentially short memory approximations
to the long memory processes; infinite moving averages were truncated to
finite ones. Thus this third region could be eliminated by careful synthesis,
i.e.\ by making the running averages much longer than the ensemble length.
Furthermore the transient region was later shown \citep{Taqqu_1970} to be
largely a feature of a programming error.

\citet{Mandelbrot_1969b} applied their $R/S$ method to many of the same data
types as \citet{Hurst_1951,Hurst_1956} and \citet{Hurst_1965}, and similarly
found significant evidence in favour of the long memory hypothesis. 
In a comparison of Hurst's $K$ with their $H$,
\citeauthor{Mandelbrot_1969b} pointed out that $K$ will tend to under-estimate
$h$ when $h>0.72$ but \emph{over}-estimate when $h<0.72$. So Hurst's
celebrated finding of a global average of 0.72 was heavily influenced by his
poor method, and his estimated standard deviation about this mean was
underestimated. This important point, that the original empirical findings
which helped spawn the subject of long memory were systematically flawed, has
long been forgotten.

Next, \citet{Mandelbrot_1969c} undertook a detailed Monte Carlo study of the
robustness to non-Gaussianity of their $R/S$ method. As previously mentioned,
in general $R/S$ was shown to be very robust. The different distributions
studied were Gaussian, lognormal, `hyperbolic' (a skewed heavy-tailed distribution --- not $\alpha$-stable but attracted to that law), and truncated
Gaussian (to achieve kurtosis lower than Gaussian). The distribution of the
un-normalised adjusted range, $R(n)$, was shown to be highly dependent on the
distribution, however the division by $S(n)$ corrected for this. For any
sequence of \iid{} random variables, their estimate $H$ was always (close to)
$1/2$.

When studying dependent cases, they considered various non-linear
transformations (such as polynomial or exponential transforms) and found that
robustness still held. However $R/S$ was shown to be susceptible in the
presence of strong periodicities; a fact rather optimistically dismissed:
``Sharp cyclic components rarely occur in natural records. One is more likely
to find mixtures of waves that have slightly different lengths \ldots''.

Finally, \citet{Mandelbrot_1969c} intriguingly replaced the Gaussian variates
in their FGN simulator with `hyperbolic' variates. Although now known have
drawbacks, this was for a long time the only attempt at simulating a
heavy-tailed long memory process.

\subsection{Reactions to Mandelbrot's model}
\label{section: reactions to Mandelbrot}

By proposing heavy tailed models to economists, Mandelbrot had had a tough
time advocating against orthodoxy \citep{Mandelbrot_2013}. Because his fractional models were
similarly unorthodox, he learned from his previous experience, and was more
careful about introducing them to hydrologists. By producing several detailed
papers covering different aspects of FBM he had covered himself against
charges of inadequate exposition. Unsurprisingly however, many hydrologists were
unwilling to accept the full implications of his papers. 

Firstly, Mandelbrot's insistence on self-similar models seemed somewhat
implausible and restrictive, and seemed to totally ignore short-term effects.
Secondly, Mandelbrot's model was continuous-time which, although necessary to
cope with self-similarity, was only useful in a theoretical context because
we live in a digital world; data are discrete and so are computers. As soon as
his models were applied to the real world they became
compromised\footnote{Mandelbrot was primarily interested in FBM; he saw the
necessary discretisation, FGN, as its derivative, both literally and
metaphorically.}: 
\begin{quotation} \singlespacing
The theory of fractional noise is
complicated by the motivating assumptions being in continuous time and the
realizable version being needed in discrete time. \citep[\S6.2]{Lawrance_1977}
\end{quotation}
In one major respect Mandelbrot was simply unlucky with timing. Soon after his
papers about FBM were published, the hugely influential book by
\citet{Box_1970} was published, revolutionising the modelling of discrete
time series in many subject areas.

Prior to 1970, multiple-lag auto-regressive or moving average models had been
used (and as previously mentioned had failed to adequately replicate the Hurst
phenomenon), but the Box--Jenkins models combined these concepts, together
with an integer differencing parameter $d$, to produce the very flexible class
of \ARIMA{p,d,q} models. As in other scientific fields, many hydrologists were
attracted to these models, and sought to explore the possibility of using them
to replicate the Hurst phenomenon.

It is important to note that ARIMA models \emph{cannot} genuinely reproduce
the \emph{asymptotic} Hurst phenomenon since all ARIMA models either have short
memory, or are non-stationary. However by choosing parameters carefully, it
can be shown that it is possible to replicate the \emph{observed} Hurst phenomenon
over a large range of $n$. \citet{OConnell_1971} was an early exponent of this
idea; specifically he used an \ARMA{1,1} model which could (roughly) preserve
a given first-lag auto-correlation as well as $h$.\footnote{For completeness, we mention that other modelling approaches were
investigated to try and replicate the Hurst phenomenon. One such model was the
so-called `broken-line' process detailed by \citet{Rodriguez_1972},
\citet{Garcia_1972}, and \citet{Mejia_1972,Mejia_1974} which sought to preserve a twice differentiable spectrum. This was criticised by \citet{Mandelbrot_1972b} and did not prosper.}

To summarise, in the early 1970s there were two distinct approaches to
modelling hydrological processes. One could use traditional AR processes (or
their more advanced ARMA cousins) which, although able to partially replicate
the Hurst phenomenon, were essentially short memory models. Alternatively one
could use Mandelbrot's FGN process in order to replicate the Hurst phenomenon
accurately. Unfortunately this dichotomy was strong and the choice of approach
largely came down to whether accounting for low- or high-frequency effects was
the principal aim for the modeller. Mandelbrot himself was well aware
\citep[c.f.][p911]{Mandelbrot_1968b} that he was suggesting switching the
usual order of priority when modelling stochastic processes. Many were
uncomfortable with this approach because, whereas the ARMA models could be
coerced into replicating the Hurst phenomenon, FGN was completely
uncustomisable with regards to high frequencies.
\begin{quotation} \singlespacing
It remains for the hydrologist to decide which type of behaviour [low- or
high-frequency] is the more important to reproduce for any particular problem.
No doubt derivations of FGN's preserving both high and low frequency effects
will eventually emerge and such a choice will not be necessary.
\citep[\S2.3]{OConnell_1971}
\end{quotation}
Further studies involving ARMA processes were undertaken by
\citet{Wallis_1973}, \cite{Lettenmaier_1977} (who proposed a mixture of an
ARMA(1,1) model with an independent AR(1) model), and the set of papers by
\citet{McLeod_1978a} and \citet{Hipel_1978b,Hipel_1978c}. These latter authors
were the first to apply to long memory processes the full Box--Jenkins
philosophy of time series estimation: model identification, parameter
estimation, and model-checking. To compare models, they were the first to use
formal procedures such as information criteria, and formally test 
residuals for whiteness. With this setup they fitted models to six long-run
geophysical time series suspected of possessing long memory, and found that in
each case the best fitting ARMA models were chosen in preference to FGN. They
also fitted more complex ARMA models (than ARMA(1,1)) and showed
again that the observed Hurst statistic can be maintained over the length of
series used.\footnote{As an aside, the set of papers by
\citeauthor{McLeod_1978a} were also remarkable for two other reasons. As
mentioned previously, they developed an \emph{exact} FGN simulator (using the
Cholesky decomposition method), which although computationally expensive,  was
the first time anyone had been able to simulate genuine long memory data.
Secondly, the authors derived a maximum likelihood estimator for the FGN
parameter $h$. This was the first proper attempt at parametric modelling of
FGN. \cite{Mandelbrot_1979} were dismissive of this approach due to the strong
assumptions needed, however from a theoretical statistical point of view it
was a clear breakthrough.}
 
Along with their practical difficulty, another ground for rejecting
Mandelbrot's models was his sweeping assertions about their physical
interpretation. Slightly paraphrasing, he claimed that, since long memory was
the only explanation for the Hurst phenomenon, the underlying physical
processes must possess long memory. This approach of inferring physics from an
empirical model was generally rejected. For a start, many were reluctant to
drop the natural Markovian assumption about nature:  
\begin{quotation} \singlespacing
The past influences the future only through its effect on the present, and thus
once a certain state of the process has been reached, it matters little for
the future development how it was arrived at. \citep{Klemes_1974}
\end{quotation} 
Indeed the renowned hydrologist Vit \citeauthor{Klemes_1974} was
a leading opponent of Mandelbrot's interpretation. As indicated earlier, he
personally suspected non-stationarity might be the true cause for the observed
Hurst phenomenon. Whilst he was convinced of the importance of the Hurst effect, and accepted FGN as an empirical model (he used the
phrase `operational model') he strongly rejected using it to gain an
understanding of physics: 
\begin{quotation} \singlespacing
An ability to simulate, and even
successfully predict, a specific phenomenon does not necessarily imply an
ability to explain it correctly. A highly successful operational model may
turn out to be totally unacceptable from the physical point of view.
\citep{Klemes_1974} 
\end{quotation} 
He likened the apparent success of Mandelbrot's FGN in explaining the Hurst
phenomenon to the detrimental effect that the Ptolemaic planetary model had
on the development of astronomy. Kleme\v{s} had strong reservations about
the concept of long memory, asking:
\begin{quotation} \singlespacing
By what sort of physical mechanism can the influence of,
say, the mean temperature of this year at a particular geographic location be
transmitted over decades and centuries? What kind of a mechanism is it that
has carried the impact of the economic crisis of the 1930s through World War
II and the boom of the 1950s all the way into our times and will carry it far
beyond? 
\end{quotation} 
though he conceded that there were in fact possible mechanisms in the man-made world, although not, in his view, in the physical one.

Kleme\v{s} was not alone in his concern over the interpretation of
Mandelbrot's models:
\begin{quotation} \singlespacing
Using self-similarity (with $h\neq 1/2$) to extrapolate the correlated
behaviour from a finite time span to an asymptotically infinite one is
physically completely unjustified. Furthermore, using self-similarity to
intrapolate [sic] to a very short time span \ldots is physically absurd.
\citep{Scheidegger_1970}
\end{quotation}
Interestingly, in his reply, \citet{Mandelbrot_1970} somewhat missed the
point:
\begin{quotation} \singlespacing
[The] self-similar model is the only model that predicts for the rescaled
range statistic \ldots precisely the same behaviour as Harold Edwin Hurst has
observed empirically. To achieve the same agreement with other models, large
numbers of \emph{ad hoc} parameters are required. Thus the model's
justification is empirical, as is ultimately the case for any model of nature.
\end{quotation} 
Yet another argument used to oppose the use of long memory models arose from a
debate about their practical value. By not incorporating long memory into
models, at how much of a disadvantage was the modeller? Clearly, this is a
context-specific question, but the pertinent question in hydrology is: by how
much does incorporating long memory into the stochastic model change the ideal
dam height? One view, shared by Mandelbrot:
\begin{quotation} \singlespacing
The preservation within synthetic sequences \ldots [of $h$] is of prime
importance to engineers since it characterizes long term storage behaviour.
The use of synthetic sequences which fail to preserve this parameter usually
leads to underestimation of long term storage requirements.
\citep{OConnell_1971}.
\end{quotation}
By ignoring the Hurst phenomenon, we would generally expect to underestimate
the ideal dam height but how quantifiable is the effect? \citet{Wallis_1972}
were the first to demonstrate explicitly that the choice of model did indeed
affect the outcome: by comparing AR(1) and FGN using the Sequential Peak
algorithm --- a deterministic method of assessing storage requirements based
on the work of \citet{Rippl_1883} and further developed in the 1960s.
\citeauthor{Wallis_1972} showed that the height depends on both the short and
long memory behaviours, and in general, FGN models require larger storage
requirements, as expected. \citet{Lettenmaier_1977} went into more detail by
looking at the distribution of the ideal dam height (rather than simply the
mean value) and found it followed extreme value theory distributions.
\citeauthor{Lettenmaier_1977} also showed that long memory inputs required
slightly more storage, thus confirming the perception that long memory models
need to be used to guard against `failure'.

However \citet{Klemes_1981} argued against using this philosophy; instead
suggesting that `failure' is not an absolute term. In the context of
hydrology, `failure' would mean being unable to provide a large enough water
supply; yet clearly a minimal deficit over a few days is a different severity
to a substantial drought over many years. Any `reasonable' economic analysis
should take this into account. \citeauthor{Klemes_1981}~claimed that the
incorporation of long memory into models used to derive the optimum storage
height is essentially a `safety factor', increasing the height by a few
percent, however ``\ldots in most practical cases this factor will be much
smaller than the accuracy with which the performance reliability can be
assessed.''

In summary therefore, Mandelbrot's work was controversial because, although it
provided an explanation of Hurst's observations, the physical interpretation
of the solution was unpalatable. There was no consensus regarding the whole
philosophy of hydrological modelling; should the Hurst phenomenon be accounted
for, and if so implicitly or explicitly? Moreover, the new concept of long
memory, borne out of the solution to the riddle, was both non-intuitive and
mathematically unappealing at the time.

Much of the debate outlined above
was confined to the hydrological community, in particular the pages of
\emph{Water Resources Research}. With the exception of articles appearing in
probability journals concerning the distributions of various quantities
related to the rescaled adjusted range, little else was known about long
memory by statisticians. This was rectified by a major
review paper by \citet{Lawrance_1977} which helped bring the attention of the
Hurst phenomenon to the wider statistical community.

One of those non-hydrologists who took up the Hurst `riddle' was the eminent
econometrician Clive Granger. In an almost-throwaway comment at the end of a
paper, \citet{Granger_1978b} floated the idea of `fractionally differencing' a
time series, whose spectrum has a pole at the origin.\footnote{It  should be pointed out that the ubiquity of $1/f$ spectra had been a puzzle to physicists since the work of  Schottky in 1918. \citet{Adenstedt_1974},   derived some properties of such processes
but his work went largely unnoticed until the late 1980s, while \cite{Barnes_1966} considered a model of $1/f$ noise explicitly based on fractional integration} Granger's
observation was followed up by both himself, and independently by the hydrologist
\citet{Hosking_1981}, who between them laid the foundations for a different
class of long memory model. This class of ARFIMA models are the most commonly
used long memory models of the present day. If the empirical findings of
\citeauthor{Hurst_1951} helped to stimulate the field, and the models of
\citeauthor{Mandelbrot_1965} helped to revolutionise the field, the class of
ARFIMA models can be said to have made the field accessible to all.

\section{Fractionally differenced models}
\label{section: fractionally differenced models}
One of the objections to Mandelbrot's
fractional Gaussian noise was that it was a discrete approximation to a
continuous process. \citet{Hosking_1981} explained how FGN can be roughly
thought of as the \emph{discrete version} of a fractional derivative of
Brownian motion. In other words, FGN is obtained by fractionally
differentiating, \emph{then} discretising. \citeauthor{Hosking_1981} proposed
to reverse this order of operations, i.e\ discretising first, then
fractionally differencing.

The advantage of this approach is that the discrete version of
Brownian motion has an intuitive interpretation; it is the simple random walk,
or ARIMA$(0,1,0)$ model. We may fractionally difference this using the
well-defined `fractional differencing operator of order $d$' to obtain the
\FAR{0,d,0} process, which for $0<d<1/2$ is stationary and possesses long
memory. From this loose derivation, we immediately see a clear advantage of
this process: it is formalisable as a simple extension to the classical
Box--Jenkins ARIMA models.

\citet{Granger_Joyeux_1980} arrived at a similar conclusion 
noticing that it was both \emph{possible} to
fractionally difference a process and, in order not to over- or under
difference data, it may be \emph{desirable} to do so. Direct motivation was
provided by \cite{Granger_1980} who showed that such processes could arise as
an aggregation of independent AR(1) processes, where the Auto-Regressive
parameters were distributed according to a Beta distribution (this aggregation
of micro-economic variables was a genuine motivation, rather than a contrived
example). Furthermore, \citeauthor{Granger_Joyeux_1980} pointed out that in
long-term forecasts it is the low frequency component that is of most
importance.\footnote{It is worth remarking that \emph{forecasting} is quite
different from \emph{synthesis} discussed earlier; the former takes an
observed sequence and, based on a statistical examination of its past,
attempts to extrapolate its future. This is a deterministic approach and,
given the same data and using the same methods, two practitioners will produce
the same forecasts. Synthesis on the other hand is a method of producing a
representative sample path of a given process and is therefore stochastic in
nature. Given the same model and parameters, two practitioners will produce
different sample paths (assuming their random number generator seeds are not
initiated to the same value). However their sequences will have the same
statistical properties.}

Both \citet{Granger_Joyeux_1980} and \citet{Hosking_1981} acknowledged that
their model was based on different underlying assumptions to Mandelbrot's
models. They also recognised the extreme usefulness of introducing long memory
to the Box--Jenkins framework. By considering their fractionally differenced
model as an ARIMA$(0,d,0)$ process, it was an obvious leap to include the
parameters $p,q$ in order to model short-term effects; thence the full
\FAR{p,d,q} model. By developing a process which could model both the short
and long memory properties, the authors had removed the forced
dichotomy between ARMA and FGN models. By being able to model both types of
memory simultaneously, ARFIMA models immediately resolved the main practical
objection to Mandelbrot's FGN model.

Depending on the individual context and viewpoint, ARFIMA models can either be
seen as pure short memory models adjusted to induce long memory behaviour, or
pure long memory models adjusted to account for short-term behaviour. ARFIMA
models are more often introduced using the former of these interpretations ---
presumably because most practitioners encounter the elementary Box--Jenkins
models before long memory --- however it is arguably more useful to consider
the latter interpretation.

Although slow to take off, the increased flexibility of ARFIMA
models, and their general ease of use compared to Mandelbrot's FGN, meant that
they gradually became the long memory model of choice in many areas including hydrology and econometrics, although we have found them still to be less well known in physics than FGN. Apart from their
discreteness (which may, or may not be a disadvantage depending on the point
of view) the only disadvantage that ARFIMA models have is that they are no
longer completely self-similar. The re-scaled partial sums of a `pure' ARFIMA$(0,d,0)$
model converge in distribution to FBM \cite[see e.g.][\S6]{Taqqu_2003}, so,
in some sense, the process can be seen as the increments of an {\em asymptotically}
self-similar process. However any non-trivial short memory component
introduces a temporal `tick' and destroys this self-similarity.

Perhaps inevitably given his original motivation for introducing self-similarity as an explanation for the Hurst phenomenon, and his further development of the whole concept of scaling into fractal theory, Mandelbrot was not attracted to ARFIMA models. Decades after their introduction, and despite their popularity, Mandelbrot would state: 
\begin{quotation} \singlespacing
[Granger] prefers a discrete-time version of FBM that differs a bit from the
Type I and Type II algorithm in \cite{Mandelbrot_1969a}. Discretization is
usually motivated by unquestionable convenience, but I view it as more than a
detail. I favor very heavily the models that possess properties of
time-invariance or scaling. In these models, no time interval is privileged by
being intrinsic. In discrete-time models, to the contrary, a privileged time
interval is imposed nonintrinsically. \citep{Mandelbrot_2002c}
\end{quotation} 
Convenience would seem to rule the roost in statistics, however, as ARFIMA-based inference is 
applied in practice far more often than FBM/FGN.  Many practitioners would argue
that it is not hard to justify 
use of a ``privileged time interval'' in a true data analysis context:
the interval at which the data are sampled and/or at which decisions based
on such data would typically be made, will always enjoy privilege in modeling
and inference. 

As we saw above, the introduction of the LRD concept into science came with Mandelbrot's application of the fractional Brownian models of Kolmogorov to an environmetric observation -- Hurst's effect in hydrology.  Nowadays, an important new environmetric application for LRD is to climate research. Here
ARFIMA plays an important role in understanding long-term climate variability and in trend estimation, but remains less well known in some user communities compared to, for example, SRD models of the Box-Jenkins type, of which AR(1) is still the most frequently applied. Conversely, in many  branches of physics the fractional $\alpha$-stable family of models including FBM remain rather better known than ARFIMA. The process of codifying the reasons for the similarities and differences between these models, and also the closely related anomalous diffusion models such as the Continuous Time Random Walk, in a way accessible to users, is underway 
but much more remains to be done here, particularly on the ``physics of FARIMA". 

\section{Conclusion}
\label{sec:conclude}

We have attempted to demonstrate the original motivation behind long memory
processes, and trace the early evolution of the concept. Debates over the nature of such processes, and their applicability or appropriateness to real life, are still ongoing. Importantly, the physical meaning of FBM has been clarified by studies which show how it plays the role of the noise term in the generalised Langevin equation when a particular (``$1/f$") choice of heat bath spectral density has been made, see for example \cite{Kupferman_2004}. Rather than draw our own conclusions,  we rather intended to  illuminate the story of this fascinating area of science, and in particular the role played by  Benoit Mandelbrot, who died in 2010. The facet of Mandelbrot's genius on show here was to use his strongly geometrical mathematical imagination to link some very arcane aspects of the theory of stochastic processes to the needs of operational environmetric  statistics. Quite how remarkable this was can only be fully appreciated when one reminds oneself of the available data and computational resources of the early 1960s, even at IBM. The wider story \citep{Mandelbrot_2008,Mandelbrot_2013} in which this paper's theme is embedded, of how he developed and applied in sequence, first the $\alpha$-stable model in economics, followed by the fractional renewal model in $1/f$ noise, and then FBM, and a fractional hyperbolic  precursor to the linear fractional stable models, and  finally a multifractal model, all in the space of about 10 years, shows both mathematical creativity and  a real willingness to listen to what the data was telling him. The fact the he (and his critics) were perhaps less willing to listen to each other is a human trait whose effects on this story-we trust-will become less significant over time.

\subsection*{Acknowledgements} 
This paper is derived from Appendix D of TG's PhD thesis \citep{graves:phd}. NWW and CF acknowledge the stimulating environment of the British Antarctic Survey, and TG and RG  that of the Cambridge Statslab, during this period. We thank Cosma Shalizi, Holger Kantz, and the participants in the International Space Science Institute programme on ``Self-Organized Criticality and Turbulence" for  discussions, and David Spiegelhalter for support. CF was supported by the German Science Foundation (DFG) through the cluster of excellence CliSAP, while NWW has recently been supported at the University of Potsdam by ONR NICOP grant N62909-15-1-N143.

\bibliography{LRD_database}
\bibliographystyle{jasa}

\end{document}